\documentclass[letterpaper,reprint,final,aps,prd,longbibliography,amsfonts,amssymb,amsmath,superscriptaddress]{revtex4-2}

\usepackage[utf8]{inputenc}
\usepackage[T1]{fontenc}
\usepackage{mathtools}
\usepackage{float}
\usepackage{hyperref}
\hypersetup{%
	pdfauthor={Giulia Maniccia and Giovanni Montani},%
	pdftitle={Quantum gravity corrections to the matter dynamics in the presence of a reference fluid}%
}

\allowdisplaybreaks[1]

\newcommand{\eu}{\mathrm{e}}
\newcommand{\iu}{\mathrm{i}}

\newcommand{\ord}[1]{\mathcal{O} \left(#1\right)}

\begin{document}
\title{Quantum gravity corrections to the matter dynamics in the presence of a reference fluid}

\author{Giulia Maniccia}
\email{giulia.maniccia@uniroma1.it}
\affiliation{Physics Department, “La Sapienza” University of Rome, P.le A. Moro 5, 00185 Roma, Italy}
\affiliation{INFN Section of Rome, “La Sapienza” University of Rome, P.le A. Moro 5, 00185 Roma, Italy}

\author{Giovanni Montani}
\email{giovanni.montani@enea.it}
\affiliation{ENEA, FNS Department, C.R. Frascati, Via E. Fermi 45, 00044 Frascati (Roma), Italy} 
\affiliation{Physics Department, “La Sapienza” University of Rome, P.le A. Moro 5, 00185 Roma, Italy}

\begin{abstract}
	We analyze the effect induced on standard quantum field theory (in a functional approach) by quantum gravity corrections to a pure classical background.
    In the framework of the Kucha\v{r} and Torre proposal for a gravity-matter theory constrained to a Gaussian reference frame, materialized as a fluid in the system evolution, we consider a Born-Oppenheimer separation of the system, regarding the gravity degrees of freedom as the slow varying component and the matter plus the Gaussian fluid as fast quantum coordinates.
    The slow gravity component obeys the Wheeler-DeWitt equation and we consider a Wentzel-Kramer-Brillouin expansion of its quantum dynamics via a Planckian parameter. The main issue of the proposed scenario is that, on one hand, we recover a modified quantum field theory in the presence of a Hermitian Hamiltonian (not affected by nonunitarity as in other approaches) and, on the other hand, we get the Gaussian fluid as a physical clock for such amended quantum theory (verifying the correct energy conditions). We also show its equivalence with the kinematical action method, used in a previous work, in the homogeneous setting.
    Then, we implement the proposed paradigm to describe the dynamics of a homogeneous free massless scalar field, living on an isotropic universe, in the presence of a cosmological constant. We completely solve the dynamics up to the first order correction in the Planckian parameter to the standard quantum field theory. We determine the explicit form of the modified scalar field wave function, due to quantum features of the cosmic scale factor evolution. Phenomenological considerations and discussions are provided.

\end{abstract}

\maketitle

\section{Introduction\label{sec:Intro}}

    One of the most puzzling questions affecting the canonical quantization \cite{bib:dirac-lecturesOnQuantumMechanics} of the gravitational field \cite{bib:dewitt1-1967,bib:dewitt2-1967,bib:dewitt3-1967,bib:kuchar-1981,bib:cianfrani-canonicalQuantumGravity,bib:montani-primordialcosmology}  is the so-called  ``frozen formalism'', i.e., the absence of a time evolution of the Universe wave function \cite{bib:rovelli-1991-time,bib:isham-1992,bib:kiefer-1994-review,bib:montani-2004}.
    This basic problem of the quantum gravitational field dynamics has been faced in the literature by many approaches, some dealing with the Schr\"{o}dinger equation \cite{bib:kuchar-1991,bib:brown-1995,bib:thiemann-2006,bib:montanizonetti-2008,bib:giulini-2012,bib:goltsev-2021,bib:dimakis-2021}, and others facing related topics \cite{bib:rovelli-1991-quantrefsyst,bib:rovelli-1991-incompatibility,bib:thiemanngiesel-2007,bib:giesel-2009,bib:chataignier-2020,bib:partouche-2021}, although this problem still remains an open issue, especially for what concerns the definition of a causality relation.

    In general, introducing a time variable for canonical quantum gravity is a troublesome task. Many approaches choose whether to follow a path of reduced quantization, i.e. defining a clock variable before the quantization procedure, or to implement the Dirac method, where the quantum version of the constraints is imposed on the system and the time parameter is identified afterward (for example, see discussions in \cite{bib:gielen-2021-frozen},\cite{bib:bojowald-2018} and references within); in this work we will focus on the second procedure. 
 
    A recurring proposal has been to somehow define time via fluid variables into the theory, such as in \cite{bib:kuchar-1991,bib:brown-1995,bib:montanizonetti-2008,bib:rotondo-2020,bib:gielen-2021-unitarity}. In this context, a very interesting attempt must be considered in the analysis in \cite{bib:kuchar-1991}, where the request to constrain the canonical procedure to a specific Gaussian reference frame has been pursued in a covariant formulation. The very suggestive result of this proposal is that the fixing procedure of a Gaussian frame has the effect to materialize such a reference frame into the dynamics, in the form of a fluid having zero pressure and thermal conductivity; actually, it reduces to simply an incoherent dust if the constraint concerns only the unit value of the metric $0-0$-component.
    The shortcoming of the proposed approach consists of the impossibility to deal with a fluid which always satisfies the so-called ``energy condition''; i.e., its physical nature is not guaranteed (for instance, the dust could be endowed with a negative energy density).

    Here, we intend to implement this idea on the level of a Wentzel-Kramer-Brillouin (WKB) limit for the gravitational field, which allows a Born-Oppenheimer (BO) separation of the gravity-matter dynamics \cite{bib:vilenkin-1989,bib:kiefer-1991,bib:bertoni-1996,bib:venturi-2017,bib:chataignier-2021,bib:venturi-2021}. This scenario based on an expansion of the quantum dynamics via a parameter of the order of the Planck scale (or in the Planck constant) allows us to recover the quantum field theory of matter on a classical curved background (in the functional representation) at the zeroth order of expansion, while at the higher one, the Hamilton-Jacobi equation for the gravitational field naturally emerges. However, in the next order of approximation, the corrections to the functional Schr\"{o}dinger equation are associated with a nonunitary contribution.

    In \cite{bib:maniccia-2021}, a detailed discussion and criticism of this type of approach was presented, and a different proposal has been inferred. In fact, instead of deriving the time evolution of the matter wave function from its parametric dependence on the classical (say better quasiclassical) gravitational variables, there the so-called ``kinematical action'' was added to the dynamics, introduced in \cite{bib:kuchar-1981} to covariantize a quantum field theory on a classical background and so outlining the natural constraints of the theory. In \cite{bib:montani-2002} the kinematical action was also implemented in a fully quantum gravity scenario to provide a time evolution for the wave function, and it was also shown that, in the classical limit, it acquires the morphology of a physical fluid. 
    In \cite{bib:maniccia-2021}, the kinematical action does not appear in the classical limit of gravity, since it is associated with a matterlike variable in the BO approximation. This way, the kinematical action simply provides the proper time evolution of quantum field theory; the quantum gravity corrections are no longer affected by nonunitarity problems.

    As a first step, here we consider the general case with the full Hamiltonian formulation of gravity and matter in the presence of a ``reference fluid''. We construct the first three orders of approximation in expansion with respect to a Planckian parameter; the analysis is developed within a Born-Oppenheimer scheme in which the gravitational degrees of freedom correspond to the slow component, while the matter sources and reference fluid are described via fast variables. It is important to stress, in this respect, that the BO separation is always performed requiring that, as in \cite{bib:vilenkin-1989}, the gravitational degrees of freedom obey a Wheeler-DeWitt equation alone (before the WKB approximation is implemented). 
    We show the general form of the quantum corrections to the functional Schr\"{o}dinger equation describing the standard quantum field theory. A key point is that the reference fluid, while contributing to Einstein's equations with its stress-energy tensor, does not appear at the first order of expansion (i.e. at the Planckian scale), where the Hamilton-Jacobi equation for the gravitational field alone is recovered. This feature must be taken into account when discussing the energy conditions, as done in \cite{bib:kuchar-1991}, since the fluid itself is not a source for gravity at the first order.
    Furthermore, we show that there exists a connection between the reference fluid procedure and the kinematical action path presented previously in \cite{bib:maniccia-2021}: indeed, when one implements only the Gaussian time condition (i.e. the reference fluid is an incoherent dust, not transporting heat), the resulting dynamics for the matter sector corresponds exactly to the results obtained via the kinematical action insertion in the homogeneous case, when the component $N^i$ of the deformation vector is set to zero.

    Then, we perform a cosmological study of the isotropic universe in the presence of a homogeneous free massless scalar field and a cosmological constant term (these last two ingredients mimic the inflation paradigm of the early Universe), and we include the fixing of a Gaussian frame into the quantization procedure.
    According to the general scheme mentioned above, the (materialized) reference frame is described by matterlike variables, and the BO separation is implemented, in order to calculate the correction to the Hamiltonian spectrum of the free scalar field, as the effect of the quantum corrections to the isotropic universe dynamics. 
    
    The main result of the present study is to recover quantum gravity corrections in the form of a Hermitian modified Hamiltonian for the scalar field. Furthermore, different from the analyses in \cite{bib:kiefer-1991,bib:kiefer-2018,bib:venturi-2020}, the matter wave function is sensitive to the quantum gravity nature of the scale factor, although its dependence is very weak, according to the series expansion in the Planck scale.
    This result offers a promising scenario to investigate more specific questions, like the deformation that quantum gravity corrections can cause on the spectrum of perturbations induced by the inflationary scenario.
    We conclude by observing that, in the proposed scheme, the idea of \cite{bib:kuchar-1991} finds a new perspective: the materialized reference frame is involved in the ``fast'' matter dynamics, and its energy density no longer enters the Hamilton-Jacobi equation, so that no violation of physical conditions takes place.
    
    The paper is organized as follows. In Sec. \ref{sec:KT}, we briefly review the reference frame fixing proposal, illustrating the main concerns regarding the Gaussian fluid (physical or nonphysical) properties.  In Sec. \ref{sec:WKBgeneral}, we introduce the reference fluid model and perform its expansion via the WKB procedure, showing the analogy with the kinematical action insertion proposed in \cite{bib:maniccia-2021} and demonstrating that the homogeneous (mini-super-space) case corresponds to the choice of a fluid without heat conduction. In Sec. \ref{sec:minisuperspaceg00} we implement the procedure for the homogeneous case, imposing only the time condition for the reference frame. Section \ref{sec:conclusions} contains the physical considerations and concluding remarks.

\section{The Kucha\v{r}-Torre model}\label{sec:KT}

    Here we briefly present the procedure discussed in \cite{bib:kuchar-1991} to develop a quantum formulation of geometrodynamics with the reference frame fixing approach and its critical issues.

    The starting point is the notion of reference fluid: in order to identify the dynamically significant components of the metric that describe the evolution (essentially, space points and a clock), one can fix a certain reference frame, which will emerge as a fluid in the system.
    However, the coordinate conditions are imposed before the constraints of vanishing super-Hamiltonian and supermomentum; in this way, the reference fluid will break the diffeomorphism invariance, requiring a reparametrization procedure to make the system covariant.

    In the original paper \cite{bib:kuchar-1991}, the reference frame chosen is the Gaussian one; however, one could, in general, construct an analogous procedure for a different system of coordinates (see discussion in \cite{bib:ishamkuchar-1985}).

    Choosing the Gaussian coordinates $X^{\mu} = (T, X^i)$, the conditions to be imposed are
    \begin{equation}
	    \label{eq:KTcondgauss}
	    \gamma^{00} = 1, \quad \gamma^{0i} = 0  ,
    \end{equation}
    being $\gamma_{\alpha \beta}$ the space-time metric (we have adopted the signature (+,-,-,-)). We are using greek indices $\mu, \alpha, \beta...$ to indicate 4D variables, and latin indices $i, j, k...$ for 3D (spatial) objects.

    Following the original paper, here we consider only the gravitational system with the Gaussian conditions. The conditions \eqref{eq:KTcondgauss} are obtained by inserting Lagrangian multipliers $\mathcal{F}, \mathcal{F}_i$ into the total action of the system that reads
    \begin{equation}
    	S = S^{g} + S^f,
    \end{equation}
    where $S^g$ is the usual Einstein-Hilbert action, and
    \begin{equation}
    	\label{eq:KTSfluid4d}
    	S^f = \int d^4x \left[ -\frac{\sqrt{-\gamma}}{2} \mathcal{F} \left(\gamma^{00}-1 \right) + \sqrt{-\gamma}\, \mathcal{F}_i \,\gamma^{0i} \right]
    \end{equation}
    contains the Gaussian conditions. As a consequence, these additional terms emerge in the Einstein's equations as a source for gravity, thus breaking the diffeomorphism invariance. 

    A \emph{reparametrization} of the action is then needed in order to work in arbitrary coordinates, other than the starting ones (i.e., to recover covariance). This reflects the fact that the Gaussian fluid is a ``privileged'' system of coordinates, but one could, in principle, choose another arbitrary set. For this reason, the Gaussian coordinates are written as functions of arbitrary coordinates $x^{\alpha}$, with associated metric $g_{\alpha \beta}$, so that the equations will be unchanged for coordinate transformations of the $x^{\alpha}$:
    \begin{equation}
    	\label{eq:KTSparam}
    	\begin{split}
    	S^f_{par} = \int d^4x &\left[ \frac{\sqrt{-g}}{2}\, \mathcal{F} \left(g^{\alpha \beta} \partial_{\alpha}T(x) \, \partial_{\beta}T(x) -1 \right) \right. \\
    	& \left. \vphantom{\frac{\sqrt{-g}}{2}} +\sqrt{-g}\, \mathcal{F}_i \left( g^{\alpha \beta} \partial_{\alpha}T(x)\, \partial_{\beta}X^i(x) \right) \right]  .
    	\end{split}
    \end{equation}
    For compactness of notation, we avoid specifying the dependence on the $x^{\alpha}$, which will be implicitly understood.
    The reparametrization procedure shall clearly include the simplest choice, which is when the reference frame is precisely the Gaussian one. Indeed, we require that \eqref{eq:KTSparam} coincides with \eqref{eq:KTSfluid4d} when the arbitrary coordinates are chosen as the Gaussian ones: $x^{\alpha}  \equiv \delta^{\alpha}_{\mu} X^{\mu}$ (we refer to the original paper for these steps). 

    Once the parametrized fluid formulation is obtained, one can investigate the role of this object. Defining
    \begin{gather}
    	\label{eq:KTfluidvel}
    	U^{\alpha} = g^{\alpha \beta} \partial_{\beta} T \, ,\\
    	\label{eq:KTfluidheat}
    	\mathcal{F}_{\alpha} = \mathcal{F}_i \partial_{\alpha}X^i \, ,	
    \end{gather}
    it can be shown that the source term in Einstein's equations has the following form:
    \begin{equation}
    	\label{eq:KTstressenergytensor}
    	T^{\alpha \beta} = \mathcal{F} U^{\alpha} U^{\beta} + \frac{1}{2} \left( \mathcal{F}^{\alpha}U^{\beta} + \mathcal{F}^{\beta} U^{\alpha} \right) .
    \end{equation}
    Here, the stress-energy tensor has no stress term; thus, it corresponds to a heat-conducting dust, with four-velocity $U^{\alpha}$, energy density $\mathcal{F}$, and heat flow $\mathcal{F}_{\alpha}$.
    Implementing only the Gaussian time condition in \eqref{eq:KTcondgauss}, but not the spatial one, the authors demonstrate that the model reduces to an incoherent dust, with only the quadratic term in $U^{\alpha}$ appearing in \eqref{eq:KTstressenergytensor}, since the fluid does not transport heat.

    The Hamiltonian description of the fluid is then computed starting from \eqref{eq:KTSparam}, implementing the 3+1 Arnowitt-Deser-Misner (ADM) foliation \cite{bib:arnowitt-1962}, with $h_{ij}$ the 3D induced metric, finding
    \begin{gather}
        \label{eq:KTfluidH}
        H^f = W^{-1} P + W W^k P_k \, ,\\
        \label{eq:KTfluidHi}
       H_i^f = P\, \partial_i T + P_k \, \partial_i X^k \,,
    \end{gather}
    where $P$, $P_k$ are the momenta canonically conjugate to the Gaussian coordinates $(T, X^k)$, and the coefficients
    \begin{gather}\label{eq:KTdefW}
        W \coloneqq (1- h^{jl} \partial_j T \,\partial_l T)^{-1/2} \, ,\\
        \label{eq:KTdefWk}
        W^k \coloneqq h^{jl} \partial_j T \,\partial_l X^k 
    \end{gather} 
    correspond to the spatial sector of the Gaussian restrictions in \eqref{eq:KTSparam}. It is important to notice that the fluid super-Hamiltonian \eqref{eq:KTfluidH} is linear and homogeneous in the momenta $P$, $P^k$, a feature which will be useful later on.

    It follows that the system composed of the gravitational sector and the reference fluid is subjected to the constraints:
    \begin{gather}
        \label{eq:KTsuperHconstraint}
        H^g+H^f = 0 \,, \\
        \label{eq:KTsuperHiconstraint}
        H_i^g + H_i^f = 0 \, ,
    \end{gather}
    that are, respectively, the Wheeler-DeWitt equation and the diffeomorphism constraint \cite{bib:arnowitt-1962,bib:cianfrani-canonicalQuantumGravity}, which have strongly vanishing Poisson brackets, since they do not depend on the new momenta $P, P_k$.

    The fluid momenta $P$, $P_k$, which emerge parabolically in the constraints and clearly separated from the other canonical variables, will play an important 
    role in the emerging dynamics. Indeed, for the general case of the heat-conducting Gaussian fluid, the quantum version of \eqref{eq:KTsuperHconstraint} and \eqref{eq:KTsuperHiconstraint} leads to a Schr\"{o}dinger equation
    \begin{equation}
        \label{eq:KTschrodinger}
        i\hbar\, \partial_t \Psi = \hat{\mathcal{H}} \Psi = \int_{\Sigma} d^3x \, H^g\, ,
    \end{equation}
    where
    \begin{equation}
        \label{eq:KTtime}
        \partial_t \Psi = \int_{\Sigma} d^3x \left. \frac{\delta \Psi \left(T(x),X^k(x),h^{jl}(x)\right)}{\delta T(x)} \right|_{T=t} \, .
    \end{equation}
    Here, the ADM splitting is chosen such that $t=T$, so the wave function $\Psi$ is still a functional of $X^k(x)$, $h^{jl}(x)$, but it is an ordinary function of the Gaussian time. In other words, we exactly choose the Gaussian time of the reference fluid as the time parameter for the space-time foliation. It is clear then that the Gaussian time choice has provided a meaningful clock for the system evolution, whose state is now a functional of the remaining variables. Other choices of the ADM splitting are possible, leading to analogous functional equations (see discussion in the original paper).

    Equation \eqref{eq:KTschrodinger} is equipped with the standard positive-definite conserved norm, that could be used to construct the Hilbert space of the states. However, as noted by the authors, the integrand of the norm cannot be easily interpreted as the probability density, since the fluid variables can be used only if the fluid itself is physical, i.e. it satisfies the energy conditions.
    In other words, the fluid can be used as a clock only if the associated stress-energy tensor satisfies the weak, dominant and strong energy conditions (a discussion on the general form of these conditions can be found in \cite{bib:visser-2017}). 

    For the incoherent dust case ($\mathcal{F}_{i}$=0) these are all encoded in the request:
    \begin{equation}
    	\label{eq:KTenergyrequest}
    	\mathcal{F} \geq 0 \, .
    \end{equation} 
    In this case, the authors show that, if condition \eqref{eq:KTenergyrequest} is satisfied at the beginning, it remains valid thanks to the algebra of the super-Hamiltonian and supermomentum operators. 
    However, this is not true in the general case, since the Gaussian fluid does not have a proper equation of state. Indeed, for the heat-conducting fluid, the weak energy condition is an inequality involving the multipliers $\mathcal{F}, \mathcal{F}_i$, which is not satisfied in principle due to the independent, arbitrary values of the multipliers, and it can also be violated during the dynamical evolution.
    
    In the Hamiltonian formalism, the energy conditions require $H^f \geq 0$, which using the constraint \eqref{eq:KTsuperHconstraint} becomes $H^g \leq 0$. This condition can be written as an equality using the Heaviside function $\Theta$, as discussed in the original paper; thus, it can be considered an additional constraint for the system. However, the authors show that its Poisson brackets with the super-Hamiltonian constraints do not always vanish, so they are not first class for the general heat-conducting Gaussian fluid. This property reflects the fact that for the heat-conducting fluid, the energy conditions in terms of multipliers are not preserved in the evolution. Thus, the system must be closed with the additional constraints $P_k =0$, which turn off the heat conduction (i.e. $\mathcal{F}_k=0$). It follows that, in this implementation, the quantum version of the energy conditions can be imposed in a consistent way only for the incoherent dust.
    
    These aspects prevent us from fully considering the Gaussian fluid as a quantum clock in this formulation, and indicate that a different path should be considered in order to provide a meaningful interpretation of the Schr\"odinger equation for the Universe wave function. We will see below how our approach overcomes this problem by restricting the reference fluid to be a proper clock only for the fast variables of the Born-Oppenheimer approximation.

\section{WKB expansion of the fluid approach}\label{sec:WKBgeneral}

    Here we present a possible formulation of the reference fluid that allows a physical interpretation of the system evolution.
    
    One of the main tools of the approach is the WKB procedure \cite{bib:dunham-1932,bib:landau-quantumMechanics}, which allows us to compute an approximate solution for the wave function satisfying a given Schr\"{o}dinger equation. We will not give an explicit review of this method and its (many) applications for the gravitational field here, referring the interested reader to the previous work \cite{bib:maniccia-2021} and references within. 
    
    The core idea is to identify a suitable expansion parameter, given the characteristics of the system under study, and perform a perturbative expansion. This allows us to compute, at each order, the functions whose exponential gives the wave function solving the equation, as in
    \begin{equation}
        \label{eq:expWKBgenerale}
        \Psi = \eu^{\frac{\iu}{\hbar} \left( \sigma_0 + b \sigma_1 + b^2 \sigma_2 + ... \right)},
    \end{equation}
    where $b$ is the chosen expansion parameter; in this way, the Schr\"{o}dinger equation applied to this wave function gives a set of equations, order by order, from which one can obtain the approximate solutions $\sigma_0, \sigma_1$, and so on. It is worth stressing here that the choice of expansion parameter is crucial in order identify the nature of the lowest limit of expansion: for example, choosing $b=\hbar$, clearly the zeroth order is the purely classical background of the theory.
    
    Two works have been developed first using this approach, i.e., considering a WKB expanded quantum system for gravity and matter. The first one \cite{bib:vilenkin-1989} used as an expansion parameter the Planck constant $\hbar$, thus considering a purely classical background; the variables were then naturally separated into quasiclassical and quantum ones. Going up to $\ord{\hbar}$, the author identified a suitable time parameter such that a functional Schr\"{o}dinger equation for the quantum variables emerged.
    The second one \cite{bib:kiefer-1991} (see also \cite{bib:kiefer-1993,bib:kiefer-1994-review}) used a different expansion parameter, proportional to the Einstein coefficient $\kappa$ (i.e. the gravitational scale), so that the system was separated into gravitational degrees of freedom and quantum matter variables. In this way, the lowest order describes pure gravity in vacuum, similar to a Born-Oppenheimer implementation of the gravity-matter system. With some different hypotheses, the expansion was performed up to the next order with respect to standard quantum field theory, finding the gravitational correction contributions to the Schr\"{o}dinger equation for the matter sector.
    
    \subsection{On the proposed physical point of view}
        
        In what follows, we develop  the Born-Oppenheimer separation between gravity and matter and the subsequent WKB expansion for the gravitational component, following \cite{bib:kiefer-1991}, although we deal with functional derivatives and take into account also the contribution of the supermomentum constraint for gravity and matter (also the basic notation follows that one in \cite{bib:kiefer-1991}). 
        However, it is to be remarked that, following the analysis in \cite{bib:vilenkin-1989}, we impose that the slow varying gravitational component obeys separately the corresponding (vacuum) Wheeler-DeWitt equation. This leads to a similar result as in \cite{bib:kiefer-1991}, but it is conceptually more coherent with a standard Born-Oppenheimer decomposition of the dynamics. 

        Apart from the similarity stressed above, our analysis is intrinsically very different from the proposals in \cite{bib:vilenkin-1989,bib:kiefer-1991,bib:kiefer-2018} in the way a time coordinate emerges for the functional Schr\"{o}dinger equation describing the quantum matter evolution. 
        In fact, in such proposals, the label time emerges throughout the dependence of the matter wave functional on the gravitational degrees of freedom, which play the role of parameters and, at the order of approximation $M$, are purely classical functions of the label time. 
        In other words, the time derivative of the functional $\chi$ is constructed by the sum:
        \begin{equation}
            \partial_t h\cdot \frac{\delta \chi}{\delta h} \, ;
        \end{equation}
        it is just the morphology of such a definition of the time coordinate at the ground of the nonunitary effects emerging at the order $1/M$ of the approximation scheme. 
        
        In the scenario here proposed, however, the time variable is provided by the reference fluid, emerging when the procedure derived in \cite{bib:kuchar-1991} is implemented via the Born-Oppenheimer WKB algorithm and the coordinates describing such a reference fluid are treated as fast (matterlike) variables. 
        More specifically, our time variable, or to say better, our time derivative, is  constructed as in \cite{bib:kuchar-1991}, but, in our approximation scheme, it emerges in the quantum matter dynamics only (no additional terms affect the vacuum gravitational Hamilton-Jacobi equation).

        In the proposed picture, the reference fluid plays a role very similar to the one of the so-called kinematical action: this additional term was postulated in \cite{bib:kuchar-1981} in order to make covariant a quantum field theory for matter by outlining the proper constraints, so that the canonical quantum prescription can be implemented. 
        Treating the fluid variables as a fast component permits us to reduce the presence of the fluid to an additional contribution for the quantum matter dynamics, very similar to the kinematical action and actually, the results obtained here overlap those discussed in \cite{bib:maniccia-2021}, where such a kinematical action has been added \emph{ab initio}. 
        It is, however, important to remark that, in both approaches, the possibility to deal with matter constraints similar to the ones of quantum physics on an assigned curved background, strictly relies on the development in the order parameter we perform; their morphology is therefore an intrinsic quantum manifestation, whose classical limit would make no sense in our scenario. 

        This very different methodology in constructing a clock for the quantum dynamics of matter has two advantages: (i) it allows us to avoid the dilemma of nonunitarity of the theory discussed in \cite{bib:kiefer-1991,bib:kiefer-2018}, and (ii) we can clarify how some difficulties of the original analysis in \cite{bib:kuchar-1991} are overcome when the Born-Oppenheimer separation takes place. 
        Actually, the role of a reference frame (or equivalently, of an emerging reference fluid) must be naturally regarded as similar to the one of matter: therefore, when quantum field theory is considered in the presence of quantum gravity corrections, it must be included in the set of matterlike variables. 
        This brings up the following consideration: clearly, when as in \cite{bib:kuchar-1991}, the full quantum gravity problem is considered, and gravity, matter, and the reference fluid are all on the same footing, the presence of a physical reference system (see also \cite{bib:rovelli-1991-time}) becomes nontrivial. Our point of view is that, if we take into account (like here) the role of the Planckian parameter of expansion for the full quantum gravity dynamics, then investigating the classical contribution of the reference fluid (becoming nonphysical since the energy conditions can be violated) has a limited sense. The reason is straightforward: with respect to the expansion in a Planckian parameter, the gravitational degrees of freedom approach at the highest orders of expansion a quasiclassical limit, while the matter and the reference fluid remain still in a quantum picture, i.e., the concept of classical matter must be limited as applied only to macroscopic phenomenological sources. 

    \subsection{Implementation of the model}
    
        In this model, we consider the gravitational field, together with the reference fluid and a self-interacting scalar field $\phi$ with potential $U_m(\phi)$. This field schematically represents the matter sector, which can be generalized for more scalar fields, and it assumes a key role in cosmological applications (see Sec. \ref{sec:minisuperspaceg00}).
        The action of such a model can be taken as:
        \begin{equation}
            \label{eq:StotinADM}
            \begin{split}
            S = \int dt& \int_{\Sigma} d^3x \left(\vphantom{\frac{1}{2}} \Pi^{ij} \dot{h_{ij}} +p_{\phi} \dot{\phi} - N ( \mathcal{H}^g +\mathcal{H}^m) \right.\\
            &\left.\vphantom{\frac{1}{2}} - N^i (\mathcal{H}_i^g+\mathcal{H}_i^m) \right)  + S^f,
            \end{split}
        \end{equation}
        where we have performed the ADM foliation \cite{bib:arnowitt-1962}, so $h_{ij}$ is the metric induced on the 3D hypersurfaces $\Sigma$, and the added term $S^f$ represents the parametrized fluid of the previous section \eqref{eq:KTSparam}. This term can be written explicitly in ADM coordinates by observing that the Gaussian reference frame conditions give the following requirements on the components of the deformation vector $N^{\mu}$:
        \begin{equation}
        	\label{eq:KTcondigaussADM}
        	N = \pm 1, \quad N^{i} = 0 
        \end{equation}
        corresponding to the fluid action
        \begin{equation}
        	\label{eq:KTStotalfluid}
    	    	S^f = \int  dt \int_{\Sigma} d^3x \sqrt{h} \left[ - \frac{\mathcal{F}}{2} \left(N-\frac{1}{N}\right) + \mathcal{F}_i N N^i \right]  .
        \end{equation}
        As done in the previous section, the fluid terms can be rewritten by using the momenta associated with the Gaussian coordinates and introducing the coefficients \eqref{eq:KTdefW} and \eqref{eq:KTdefWk}, so that the fluid super-Hamiltonian and supermomentum are \eqref{eq:KTfluidH} and \eqref{eq:KTfluidHi}.
        
        From Eq. \eqref{eq:StotinADM}, the Hamiltonian formalism of the theory is straightforward. However, even in the presence of matter fields, it has been shown \cite{bib:kuchar-1981} that the constraints describing the Wheeler-DeWitt equation and the diffeomorphism invariance are not removed; i.e., the total super-Hamiltonian and supermomentum must still vanish:
        \begin{gather}
            \label{eq:WKBHconstr}
            H = H^f + H^g + H^m = 0 \, ,\\
            \label{eq:WKBHiconstr}
            H_i = H^f_i + H^g_i + H^m_i = 0 \, .
        \end{gather}
        Here, the reference fluid functions have already been specified in \eqref{eq:KTfluidH} and \eqref{eq:KTfluidHi}. The gravitational components in \eqref{eq:WKBHconstr} and \eqref{eq:WKBHiconstr} are easily computed \cite{bib:arnowitt-1962}:
        \begin{gather}
            \label{eq:Hgrav}
            H^g = -\frac{\hbar^2}{2M} \left( \nabla_g^2 + g\cdot \nabla_g \right) + M \, V \, , \\
            \label{eq:Higrav}
            H_i^g = \vphantom{\frac{1}{2}} 2 \iu \hbar\, h_i\, D\cdot \nabla_g \, ,
        \end{gather}
        where $\nabla_g$ indicates the derivatives with respect to the metric variables $h_{ij}$, the term $g
        \cdot \nabla_g = g_{ij} \frac{\delta}{\delta h_{ij}}$ is inserted to account for generic factor orderings \cite{bib:kiefer-1991}, and in the supermomentum we write $h_i\, D\cdot \nabla_g = h_{ij} D_k \frac{\delta}{\delta h_{kj}}$.
        
        Finally, the scalar matter field components in \eqref{eq:WKBHconstr} and \eqref{eq:WKBHiconstr} are
        \begin{gather}
            \label{eq:WKBmatterH}
            H^m = -\hbar^2 \nabla_m^2 +U_m \, ,\\
            \label{eq:WKBmatterHi}
            H_i^m = - (\partial_i \phi) \nabla_m \, ,
        \end{gather}
        with $\nabla_m^2 = \frac{1}{2\sqrt{h}} \frac{\delta^2}{\delta \phi^2}$, and $U_m$ includes the spatial gradients.
    
        In Eqs. \eqref{eq:Hgrav} and \eqref{eq:Higrav}, we have rewritten the Einstein coefficient in terms of the following parameter:
        \begin{equation}
            \label{eq:defM}
            M \equiv \frac{c^2}{32\pi G} = \frac{c m_{Pl}^2}{4\hbar} ,
        \end{equation}
        which is directly linked to the square of the reduced Planck mass $m_{Pl}$. We will choose this parameter for the WKB expansion, since it allows for a clear separation between the gravitational background and the other components.
        
        Following a scheme that is similar to the BO separation \cite{bib:born-1927} (as done previously in \cite{bib:vilenkin-1989,bib:kiefer-1991,bib:maniccia-2021}), we postulate that the system can be separated into a slow quantum gravitational sector and a fast quantum component including the reference fluid and the matter field, such that the wave function can be taken as
        \begin{equation}
    	    \label{eq:startingpsi}
    	    \Psi \left(h_{ij}, \phi, X^{\mu} \right) = \psi \left(h_{ij}\right) \chi \left( \phi, X^{\mu} ; h_{ij}\right),
        \end{equation}
        where $X^{\mu}$ are the Gaussian coordinates, $\psi$ is the function associated with the slow gravitational background, and $\chi$ is the function for the fast matter sector (depending parametrically on the background metric). Following the WKB method, we can rewrite \eqref{eq:startingpsi} expanding both the gravity function $\psi$ and the matter function $\chi$:
        \begin{equation}
            \label{eq:psiinizialeWKB}
            \Psi\left(h_{ij}, \phi, X^{\mu} \right) = \eu^{ \frac{\iu}{\hbar} \left(M S_0 +S_1 + \frac{1}{M} S_2\right) } \eu^{\frac{\iu}{\hbar} \left(Q_1 + \frac{1}{M} Q_2\right) } \, .
        \end{equation}
        Here, the expansion is performed up to order $1/M$, sufficient for the purpose of this paper. In this notation, the functions $Q_n (\phi, X^{\mu} ; h_{ij})$ are associated with the fast matter sector (i.e., the first contributions to the matter function $\chi$), and similarly, $S_n (h_{ij})$ are the slow background functions.
        
        To follow the BO approach, the adiabatic condition must also be implemented, stating that the functional gradients of the fast wave function with respect to slow coordinates are small:
        \begin{equation}
            \label{eq:condizderivate}
            \frac{\delta Q_n}{\delta h_{ij}} = \ord{\frac{1}{M}},
        \end{equation}
        which must be implemented order by order to the fast functions $Q_n$.
        We also require that:
        \begin{equation}
            \label{eq:condizHchi}
            \frac{\hat{H}^m \chi}{\hat{H}^g \chi} = \ord{\frac{1}{M}},
        \end{equation}
        which states that the fast matter sector lives at a smaller energy scale with respect to gravity, i.e. it is of a smaller order in the expansion parameter. For the same reason, we can consider the effect of the matter fields to be negligible at the Planck scale, so that the gravitational wave function satisfies a set of constraints on its own:
        \begin{gather}
            \label{eq:vincologravH}
            \hat{H}^g \,\psi (h_{ij}) =0 \, , \\
            \label{eq:vincolo gravHi}
            \hat{H}_i^g \,\psi(h_{ij}) =0 \, .
        \end{gather}
            
        These equations are to be adjoined to the total constraints of the system deriving from \eqref{eq:WKBHconstr} and \eqref{eq:WKBHiconstr}; namely, the system of equations to be expanded order by order reads explicitly:
        \begin{gather}
            \label{eq:Hgravconstrlong}
            \left[-\frac{\hbar^2}{2M} \left( \nabla_g^2 + g\cdot \nabla_g \right) + M V \right] \psi = 0 \, , \\
            \label{eq:Higravconstrlong}
            \left[\vphantom{\frac{1}{2}} 2\iu \hbar\, h_i\, D\cdot \nabla_g \right] \psi =0 \, ,\\
            \label{eq:Htotconstrlong}
            \begin{split}
               \left[ -\frac{\hbar^2}{2M} \right.& \left( \nabla_g^2 + g\cdot \nabla_g \right) + M V -\hbar^2 \nabla_m^2 +U_m\\
                &\left. \vphantom{\frac{1}{2}}+ W^{-1} P + W W^k P_k \right] \Psi =0 \, ,
            \end{split}\\
            \label{eq:Hitotconstrlong}
            \begin{split}
               \left[ \vphantom{\frac{1}{2}} 2 \iu \hbar \right. & h_i \,D\cdot \nabla_g - (\partial_i \phi) \nabla_m +P\, \partial_i T \\
               &\left. \vphantom{\frac{1}{2}}+ P_k \, \partial_i X^k \right] \Psi =0 \, .
            \end{split}
        \end{gather}
   
        The zeroth order is $\ord{M}$, where one obtains
        \begin{subequations}
        \begin{gather}
            \frac{1}{2} (\nabla_g S_0)^2 +V =0 \, ,\\
            -2 h\, D\cdot \nabla_g S_0 =0 
        \end{gather}
        \end{subequations}
        corresponding to the Hamilton-Jacobi equation for gravity, i.e., in vacuum, and to the diffeomorphism invariance for $S_0$.
        
        At the next order $\ord{M^0}$, the equations give
        \begin{subequations}
        \begin{gather}
            \label{eq:HgravOrdM0}
            -\iu\hbar \nabla_g^2 S_0 +2(\nabla_g S_0)(\nabla_g S_1) +\iu\hbar \,g\cdot \nabla_g S_0 =0 \vphantom{\frac{1}{2}} \, ,\\
            \label{eq:HigravOrdM0}
            -2h_i\, D\cdot \nabla_g S_1 =0 \vphantom{\frac{1}{2}}\, ,\\
            \label{eq:HtotOrdM0}
            \begin{split}
                -\iu\hbar &\nabla_g^2 S_0 +2(\nabla_g S_0)(\nabla_g S_1) +\iu\hbar \,g\cdot \nabla_g S_0 \\
                &-2\iu\hbar \nabla_m^2 Q_1 +U_m -W^{-1} \frac{\delta Q_1}{\delta T}\\
                &-W W^k \frac{\delta Q_1}{\delta X^k} =0 \,,
            \end{split}\\
            \label{eq:HitotOrdM0}
            \begin{split}
            -2h_i\, &D\cdot \nabla_g S_1 -\frac{\iu}{\hbar} (\partial_i \phi) \nabla_m Q_1 - (\partial_i T) \frac{\delta Q_1}{\delta T} \\
            &- (\partial_i X^k) \frac{\delta Q_1}{\delta X^k} =0 \, .
            \end{split}
        \end{gather}
        \end{subequations}
        Here, the gravitational constraints simplify Eqs. \eqref{eq:HtotOrdM0} and \eqref{eq:HitotOrdM0}. Remembering that, at this order, the quantum matter wave function is
        \begin{equation}
            \label{eq:defchi0}
            \chi_0 = \eu^{\frac{\iu}{\hbar} Q_1} ,
        \end{equation}
        Eqs. \eqref{eq:HtotOrdM0} and \eqref{eq:HitotOrdM0} can be combined in the following:
        \begin{equation}
        \begin{split}
            \label{eq:schrodchi0}
            \mathcal{H}^m \chi_0& = \int d^3x (N H^m +N^i H^m_i) \\
            & = \int d^3x \left[ N\left( W^{-1} \frac{\delta}{\delta T} +W W^k \frac{\delta}{\delta X^k} \right) \right.\\
            & \left. +N^i\left( (\partial_i T)\frac{\delta}{\delta T} +(\partial_i X^k) \frac{\delta}{\delta X^k} \right) \right] \chi_0 \\
            &= \iu \hbar \frac{\delta}{\delta \tau} \chi_0 \,,
        \end{split}
        \end{equation}
        which is a functional equation, with the same form as the Schr\"{o}dinger one, when one defines the quantum clock of the theory with the time derivative as above. We remark that definition \eqref{eq:schrodchi0} is a generalization of the time derivative implemented in the Kucha\v{r}-Torre model when choosing the time parameter as exactly the Gaussian time \eqref{eq:KTtime}. However, here we maintain the Gaussian coordinates as functions of the generalized parameters; thus, we do not need to implement a specific coordinate choice with this definition.  We also point out that it is possible to implement a simplified form of the time derivative by using the supermomentum constraint \eqref{eq:WKBHiconstr}, since it takes away a term from the right-hand side.
        
        Going up to the next order $\ord{M^{-1}}$ one finds
        \begin{subequations}
        \begin{gather}
            \label{eq:HgravOrdM-1}
            \begin{split}
                -\iu \hbar \nabla_g^2 &S_1 + (\nabla_g S_1)^2 +2(\nabla_g S_0)(\nabla_g S_2) \\
                &+\iu \hbar \,g\cdot \nabla_g S_1 =0 \, ,
            \end{split}\\
            \label{eq:HigravOrdM-1}
            -2 h_i D\cdot \nabla_g S_2 =0 \vphantom{\frac{1}{2}}\, ,\\
            \label{eq:HtotOrdM-1}
            \begin{split}
               -\frac{\iu \hbar}{2} \nabla_g^2 &S_1 + \frac{1}{2} \left( \vphantom{\frac{1}{2}}(\nabla_g S_1)^2 +2(\nabla_g S_0)(\nabla_g S_2) \right. \\
                &\left.\vphantom{\frac{1}{2}} +2M (\nabla_g S_0)(\nabla_g Q_1) \right) +\frac{\iu \hbar}{2} g\cdot \nabla_g S_1 \\
                &-\iu \hbar \nabla_m^2 Q_2 +2(\nabla_m Q_1)(\nabla_m Q_2) \\
                &- W^{-1} \frac{\delta Q_2}{\delta T} - W W^k \frac{\delta Q_2}{\delta X^k} =0 \, ,
            \end{split}\\
            \label{eq:HitotOrdM-1}
            \begin{split}
               -2 h_i D\cdot &\left( \nabla_g S_2 +M \nabla_g Q_1 \right) -\frac{\iu}{\hbar} (\partial_i \phi) \nabla_m Q_2\\
               &-(\partial_i T) \frac{\delta Q_2}{\delta T} -(\partial_i X^k) \frac{\delta Q_2}{\delta X^k} =0 \, .
            \end{split}
        \end{gather}
        \end{subequations}

        Here again, the first terms in \eqref{eq:HtotOrdM-1} and \eqref{eq:HitotOrdM-1} disappear due to the gravitational constraints, leaving terms that contain the functions $Q_2$ and $Q_1$. However, noting that the quantum matter wave function at $\ord{M^{-1}}$ is
        \begin{equation}
            \label{eq:defchi1}
            \chi_1 = \eu^{\frac{\iu}{\hbar} \left( Q_1 +\frac{1}{M} Q_2 \right)},
        \end{equation}
        and making use of the adiabatic condition \eqref{eq:condizderivate}, we can sum Eqs. \eqref{eq:HtotOrdM-1} and \eqref{eq:HitotOrdM-1} (of order $M^{-1}$) with those found at the previous order for $Q_1$ (of order $M^0$). Therefore, we find from the super-Hamiltonian constraints,
        \begin{equation}
        \label{eq:superHconQ1Q2}
            \begin{split}
            (\nabla_g &S_0)(\nabla_g Q_1) -\iu \hbar \left( \nabla_m^2 Q_1 +\frac{1}{M} \nabla_m^2 Q_2 \right) \\
            &+(\nabla_m Q_1)^2 +\frac{2}{M}(\nabla_m Q_1)(\nabla_m Q_2) + U_m \\
            &-\left( W^{-1} \frac{\delta}{\delta T} + W W^k \frac{\delta}{\delta X^k} \right) \left( Q_1 +\frac{1}{M} Q_2\right) = 0\, ,
            \end{split}
        \end{equation}
        where we obtain the matter super-Hamiltonian $H^m$ \eqref{eq:WKBmatterH} applied to $\chi_1$ plus an extra term, and the terms with the functional derivatives with respect to the Gaussian coordinates.
        
        From the supermomentum constraint, with the same procedure, one finds
        \begin{equation}
        \label{eq:superHiconQ1Q2}
            \begin{split}
                -2h_i &D\cdot \nabla_g Q_1 -\frac{\iu}{\hbar} (\partial_i \phi) \left( \nabla_m Q_1 +\frac{1}{M} \nabla_m Q_2 \right)\\
                &-(\partial_i T) \left( \frac{\delta Q_1}{\delta T} +\frac{1}{M} \frac{\delta Q_2}{\delta T} \right) \\
                &-(\partial_i X^k) \left( \frac{\delta Q_1}{\delta X^k} +\frac{1}{M} \frac{\delta Q_2}{\delta X^k} \right) =0 \, ,
            \end{split}
        \end{equation}
        which again reconstructs the matter supermomentum $H^m_i$ \eqref{eq:WKBmatterHi}, plus an extra term and the ones with the functional derivatives.
        
        Now, in order to reobtain the time derivative defined in \eqref{eq:schrodchi0}, we multiply \eqref{eq:superHconQ1Q2} by $N$ and \eqref{eq:superHiconQ1Q2} by $N^i$, summing them and integrating over the spatial hypersurfaces $\Sigma$:
        \begin{equation}
        \label{eq:schrodfin}
            \begin{split}
                \iu\hbar \frac{\delta}{\delta \tau} & \chi_1 = \int d^3x \left[ N \left( W^{-1} \frac{\delta}{\delta T} +W W^k \frac{\delta}{\delta X^k} \right) \right.\\
                &\left. + N^i \left( (\partial_i T) \frac{\delta}{\delta T} +(\partial_i X^k) \frac{\delta}{\delta X^k} \right) \right] \chi_1 \\
                =& \mathcal{H}^m \chi_1 + \int d^3x \left[ N\left( -\iu\hbar \nabla_g S_0 \cdot \nabla_g \right) + \right.\\
                &\left. N^i \left( 2\iu\hbar h_i D\cdot \nabla_g \right) \right] \chi_1\,.
            \end{split}
        \end{equation}
        
        It is evident that the quantum matter dynamics at $\ord{M^{-1}}$ is modified by the terms in \eqref{eq:schrodfin}, which are due to the slow quantum gravitational background; therefore, they are quantum gravity contributions.
        
        It is important to stress that these corrective terms are isomorphic to the ones found with the kinematical action procedure in \cite{bib:maniccia-2021}. In that work, it was also shown that they are indeed unitary; thus, the nonunitarity problem is overcome with this approach. The kinematical action can be then thought of as a reference frame, which (once fixed) emerges in the formalism as a fluid with the properties discussed above.
        
        Another important point is the correspondence between the incoherent dust case and the kinematical action homogeneous case. In fact, if one takes $\mathcal{F}_i =0$ (i.e. a fluid with null heat conductivity, as seen before), the equations greatly simplify (for details, see \cite{bib:kuchar-1991}). The remarkable result is that the remaining corrective terms in \eqref{eq:schrodfin} exactly mimic the ones found in \cite{bib:maniccia-2021}, by adding the kinematical action for the gravity-matter system in the homogeneous setting (which, in that formalism, corresponds to the condition $N^i=0$ and the supermomentum constraints are identically satisfied). This property can be justified with the fact that there is a correlation between the two approaches, since the kinematical action was added exactly to play the role of a reference system in the previous work.

\section{Mini-super-space dynamics with \texorpdfstring{$g^{00}=1$}{ } }\label{sec:minisuperspaceg00}
    In this section, we show a simple cosmological application of the procedure previously analyzed, choosing a model for the universe with suitable characteristics in order to mimic a slow-roll inflation period. We select an isotropic Universe, with a free inflaton field and a cosmological constant that accounts for the almost constant inflaton potential. Evidently, due to the requirement of an isotropic model, the spatial term of the Gaussian coordinates vanishes identically and the reference time coincides with Gaussian time.
    
    In order to deal with a gravity-matter Lagrangian as restricted to a reference frame having $g^{00}=1$, we must suitably add a corresponding constraint to the total action. 
    If we denote by $T$ the time variable associated with the fixed reference system (i.e., the Gaussian fluid), the constraint to be imposed covariantly reads
    \begin{equation}
    	g^{\mu\nu}\partial_{\mu}T\partial_{\nu}T - 1 = 0 \, , 
    	\label{Gaussconstraint}
    \end{equation}
    where we set to unity the speed of light and $\mu ,\nu =0,1,2,3$.
    Thus, the total action reads as follows \cite{bib:kuchar-1991}:
    \begin{equation}
    	\begin{split}
    	S = \int d^4x\sqrt{-g} &\left\{ -\frac{1}{2\kappa}\left( R + 2\Lambda\right) + \frac{1}{2}g^{\mu\nu}\partial_{\mu}\phi\, \partial_{\nu}\phi  \right.\\
    	&\left. +\frac{\mathcal{F}}{2}\left(g^{\mu\nu}\partial_{\mu}T\, \partial_{\nu}T - 1\right)\right\} \, , 
    	\label{startingS}
    	\end{split}
    \end{equation}
    where $R$ denotes the Ricci scalar, while $\kappa$ is the Einstein constant. As explained before, $\mathcal{F}$ is a Lagrangian multiplier and its variation leads to the vanishing behavior of the last Lagrangian term (responsible for the reference frame fixing). 

    Let us now consider a flat Robertson-Walker (isotropic) universe; i.e., we deal with the ADM line element \cite{bib:montani-primordialcosmology}
    \begin{equation}
    	ds^2 = N(t)^2dt^2 - a(t)^2 \left(dx^2 + dy^2 + dz^2\right) \, , 
    	\label{RWmetric}
    \end{equation}
    where $N$ is the lapse function and $a$ the cosmic scale factor (we use $c=1$). The associated Ricci scalar is:
    \begin{equation}
        \label{eq:applicR}
        R = 6 \left( \frac{\Ddot{a}}{a} +\frac{\dot{a}^2}{a^2} \right) .
    \end{equation}
    Taking equal to unity the fiducial volume over which the spatial integration is performed, and observing that the homogeneity of the model implies $\phi = \phi(t)$, $T=T(t)$, and $\mathcal{F}=\mathcal{F}(t)$ , the action (\eqref{startingS}) reads as 
    \begin{equation}
    	\begin{split}
    	S_{RW} = \int dt &\left\{ 
    	-\frac{6}{\kappa}\left( 
    	\frac{a \, \dot{a}^2}{2N} + 
    	\frac{N\Lambda a^3}{6}\right) + \frac{a^3\dot{\phi}^2}{2N} \right.\\
    	& \left. +\frac{Ma^3}{2} \left( \frac{\dot{T}^2}{N} - N\right)\right\}
    	\, , 
    	\label{SinADMvar}
    	\end{split}
    \end{equation}
    where the dot denotes differentiation with respect to the time variable $t$. The spatial component $N^i$, and so the supermomentum functions $H_i$, are not present since the supermomentum constraint is automatically satisfied due to symmetry of the model.

    Since the Lagrangian term corresponding to the reference frame fixing vanishes identically, its Hamiltonian contribution is only $p_T\dot{T}$, $p_T$ being the conjugate momentum to the variable $T$ (coinciding with the synchronous time variable). 
    Furthermore, the momentum $p_T$ results to be defined as 
    \begin{equation}
    	p_T = \mathcal{F} a^3\frac{\dot{T}}{N}
    	\, . 
    	\label{pTdef}
    \end{equation}
    Hence to ensure the right relation $\dot{T} = N$, we have to require $p_T = \mathcal{F} a^3$. 

    Finally, it is easy to check that the total action \eqref{SinADMvar} rewrites in the Hamiltonian formulation as
    \begin{equation}
    	S_{RW} = \int dt \left\{ p_a\dot{a} + p_{\phi}\dot{\phi} + p_T\dot{T} - NH \right\} 
    	\label{eq:SwithH}
    \end{equation}
    with
    \begin{equation}
    	H \equiv -\frac{\kappa}{12}\frac{p_a^2}{a} + \frac{\Lambda}{\kappa} a^3 + \frac{p_{\phi}^2}{2a^3} + p_T	\, , 
    	\label{eq:Hinmomenta}
    \end{equation}
    where $p_a$ and $p_{\phi}$ denote the conjugate momenta to $a$ and $\phi$, respectively (in the mini-super-space, the functional dependence and derivatives reduce to the simple functions and partial derivatives only). 
    
    Using the definition \eqref{eq:defM} to explicitly show the expansion parameter, the Wheeler-DeWitt constraint \eqref{eq:KTsuperHconstraint} with the obtained Hamiltonian [which coincides with $N H$,  where $H$ is \eqref{eq:Hinmomenta}, up to the fiducial volume set to unit], clearly translates to
    \begin{equation}
        \label{eq:applicationWDW}
        \left( \frac{\hbar^2}{48 M a} \partial_a^2  +4M \Lambda a^3 -\frac{\hbar^2}{2a^3} \partial_{\phi}^2  -i\hbar \, \partial_T \right) \Psi =0\,.
    \end{equation}
    Here, we have chosen without loss of generality, the natural operator ordering by setting  $g \cdot \nabla_g=0$; we have also specified the gradients using the notation $\partial_a = \frac{\partial}{\partial a}$ for derivatives.
    
    Following the same WKB expansion implemented in the previous section, we separate and expand the total wave function of the isotropic universe as in \eqref{eq:psiinizialeWKB}, identifying the functions $S_n(a)$ for the isotropic background and the functions $Q_n(T,\phi;a)$ for the matter components (scalar field and Gaussian fluid time). 

    We also require the conditions \eqref{eq:condizHchi} and \eqref{eq:condizderivate} together with the Wheeler-DeWitt constraint for the gravitational sector, which in this case reads
    \begin{equation}
    \begin{split}
        \label{eq:applicazHgrav}
        &H^g \, \psi(a) \\
        & = \left( \frac{\hbar^2}{48a M} \,\partial_a^2 +4M\Lambda a^3 \right) \eu^{\frac{\iu}{\hbar} \left(M S_0 + S_1 +\frac{1}{M}S_2 \right)} =0\, .
    \end{split}
    \end{equation}
    
    The total super-Hamiltonian constraint, taking contributions from the matter components, is explicitly
    \begin{equation}
    \label{eq:applicazHtot}
        \begin{split}
        (H^g &+H^f +H^{\phi}) \,\Psi (\phi, T;a)\\
        =&\left( \frac{\hbar^2}{48a M} \,\partial_a^2 +4M \Lambda a^3  -\frac{\hbar^2}{2a^3} \partial_{\phi}^2 \iu \hbar \,\partial_T \right) \\
        & \times \eu^{\frac{\iu}{\hbar} \left(M S_0 + S_1 +Q_1  +\frac{1}{M} (S_2+Q_2) \right)} =0  \, .
        \end{split}
    \end{equation}
    
    Proceeding order by order, we first obtain the Hamilton-Jacobi equation for the gravitational background at $\ord{M^1}$:
    \begin{equation}
        \label{eq:applicazM1grav}
        -\left( \partial_a S_0 \right)^2 + 192 \Lambda a^4 =0\,,
    \end{equation}
    which gives the solution for the classical action $S_0$.
    
    At $\ord{M^0}$, we obtain
    \begin{gather}
        \label{eq:applicazM0grav}
        \iu \hbar\partial_a^2 S_0 - 2\partial_a S_0 \partial_a S_1 = 0\, ,\\
        \label{eq:applicazM0tot}
        \begin{split}
        \frac{\iu \hbar}{48 a} &\partial_a^2 S_0 - \frac{1}{24a} \partial_a S_0 \partial_a S_1 -\frac{\iu \hbar}{2a^3} \partial_{\phi}^2 Q_1\\
        &+\frac{1}{2a^3} \left(\partial_{\phi} Q_1\right)^2 = -\partial_T Q_1 \, ,
        \end{split}
    \end{gather}
    which can be rewritten by inserting the first equation into the second one and labeling as $\chi_0 = \eu^{\frac{\iu}{\hbar}Q_1}$ the matter wave function at this order, as the following:   
    \begin{equation}
        \label{eq:applicazchi0}
        -\frac{\hbar^2}{2a^3} \partial_{\phi}^2 \chi_0 = H^m \chi_0 = \iu \hbar \, \partial_T \chi_0 \, .
    \end{equation}
    
    At $\ord{M^{-1}}$, one obtains
    \begin{gather}
       \label{eq:applicazM-1grav} 
       \iu \hbar \partial_a^2 S_1 - \left(\partial_a S_1\right)^2 -2 \partial_a S_0 \partial_a S_2 =0 \, ,\\
       \label{eq:applicazM-1tot}
       \begin{split}
           \frac{\iu \hbar}{48 a} &\partial_a^2 S_1 -\frac{1}{48a}\left( \left(\partial_a S_1\right)^2 +2 \partial_a S_0 \partial_a S_2 \right. \\
           &\left. \vphantom{\frac{1}{2}} +2M \partial_a S_0 \partial_a Q_1 \right) -\frac{\iu \hbar}{2a^3 } \partial_{\phi}^2 Q_2\\ 
           &+\frac{1}{a^3 }\partial_{\phi} Q_1 \partial_{\phi} Q_2= - \partial_T Q_2 \, .
       \end{split}
    \end{gather}
    Here again, the solution $S_2$ from the first equation simplifies the form of the second one, leaving
    \begin{equation}
        \begin{split}
            \label{eq:appliceqQ2}
            -\frac{M}{24a} &\partial_a S_0 \partial_a Q_1 -\frac{\iu \hbar}{2a^3} \partial_{\phi}^2 Q_2 +\frac{1}{a^3}\partial_{\phi} Q_1 \partial_{\phi} Q_2\\
            &= \vphantom{\frac{1}{2}} -\partial_T Q_2 \, .
        \end{split}
    \end{equation}
    
    Remembering that the matter wave function at this order is
    \begin{equation}
        \label{eq:applicazchi1}
        \chi_1 = \eu^{\frac{\iu}{\hbar} \left(Q_1 +\frac{1}{M} Q_2 \right) } ,
    \end{equation}
    and that by hypothesis \eqref{eq:condizderivate}, the term containing $\partial_a Q_2$ is of higher order in the expansion, we can write summing \eqref{eq:appliceqQ2} with \eqref{eq:applicazchi0}:
     \begin{equation}
     \label{eq:applicschrodfin}
        \left( -\frac{\hbar^2}{2a^3} \partial_{\phi}^2 - \iu \hbar \frac{M}{24a} (\partial_a S_0)  \partial_a \right) \chi_1 = \iu \hbar \, \partial_T \chi_1 \,.
     \end{equation}
     
    We observe that the additional term in the action is isomorphic to the implementation of the so-called kinematical action discussed in \cite{bib:maniccia-2021}. In fact, in the mini-super-space we have that the normal to the spatial hypersurfaces can be taken as $n^{\mu} = (1,\vec{0})$ and $\partial_t y^{\mu} \rightarrow \dot{T}$, therefore obtaining the same expressions of \eqref{eq:applicschrodfin} for the matter dynamics. 

    As discussed in \cite{bib:kuchar-1991}, the emerging fluid violates the so-called energy condition and actually its energy is not positive definite. 
    We have seen in the previous section that, if we interpret this contribution as a ``fast variable'' in the sense of a Born-Oppenheimer approximation, this nonphysical character of the emerging synchronous fluid is overcome, since it behaves like a quantum matter component,so it can play the role of a clock for the scalar field quantum dynamics.

    \subsection{Solution of the perturbative scheme} 
    We now compute an explicit solution of the mini-super-space application discussed above. 
    
    Starting from the gravitational solutions, at $\ord{M^1}$ Eq. \eqref{eq:applicazM1grav} gives
    \begin{equation}
        \label{eq:applicsolS0}
        S_0(a) = -\frac{8\sqrt{3}}{3} \sqrt{\Lambda} \left( a^3 -a_0^3 \right) ,
    \end{equation}
    where $a_0$ is the value of the cosmic scale factor at a reference time (e.g., the start of the slow-roll phase); the negative solution has been selected to correspond to an expanding universe.
    At $\ord{M^0}$, we obtain from \eqref{eq:applicazM0grav},
    \begin{equation}
        \label{eq:applicsolS1}
        S_1(a) = \iu\hbar \, \log \left(\frac{a}{a_0}\right) \, .
    \end{equation}
    Finally, at $\ord{M^{-1}}$ we get from \eqref{eq:applicazM-1grav}
    \begin{equation}
        \label{eq:applicsolS2}
        S_2(a) = -\frac{\hbar^2}{24\sqrt{3} \sqrt{\Lambda}} \left( a^{-3} - a_0^{-3} \right) .
    \end{equation}
    
    We now focus on the fast matter solution. For the computation of these functions, it is useful to work in Fourier space, using the previous notation for the conjugated momenta $p_{\phi}$ and $p_a$, so that the general solution takes the form
    \begin{equation}
        \chi_1(a, \phi, T) = \int dp_{\phi} \int dp_a \,\tilde{\chi} (p_{\phi}, p_a, T) f(p_{\phi}, p_a) \,,
    \end{equation}
    where $f$ is a generic weight function, that we will take in Gaussian form in what follows.
    
    At $\ord{M^0}$, the dynamics is described by \eqref{eq:applicazchi0}, so that the solution corresponds to the natural plane wave for quantum matter on a (classical) curved background:
    \begin{equation}
        \label{}
        \tilde{\chi_0} = e^{-\iu \hbar \frac{p_{\phi}^2}{2a^3} T} .
    \end{equation}

    The quantum gravity effects emerge at the next order, where the matter dynamics is described by Eq. \eqref{eq:applicschrodfin}. 
   To solve it, it is convenient to use a rescaled time parameter
    \begin{equation}
        \label{eq:applicaztimea3}
        d\tau = \frac{dT}{a^3} \, .
    \end{equation}
    In this way, the dynamics for the fast matter function $\chi_1$ is
    \begin{equation}
        \label{eq:appliceqchi1Fourier}
        \iu\hbar\, \partial_{\tau} \tilde{\chi}_1 = \frac{\hbar^2 p_{\phi}^2}{2}  \tilde{\chi_1} +\frac{\hbar\, p_a\, (-\tau)^{7/3}}{3 (3\Lambda)^{1/6}} \,\tilde{\chi_1} \,,
    \end{equation}
    which is solved by the corrected plane wave
    \begin{equation}
        \label{eq:applicsolchi1}
        \tilde{\chi_1} = \exp{\left( -\iu\hbar \frac{ p_{\phi}^2 }{2} \tau \,+ \iu \frac{p_a \, (-\tau)^{7/3}}{7(3\Lambda)^{1/6}}  \right)} .
    \end{equation}
   
    To understand the effects of the quantum gravity corrections, here we provide some plots for the modified wave function at $\ord{M^{-1}}$ which we computed. 
    In order to deal, in principle, with a localized probability density, we consider a starting Gaussian wave packet:
    \begin{equation}
    \begin{split}
        \label{eq:applicwavepacket}
        f (p_a, p_{\phi})=& \frac{1}{\sqrt{(2\pi)^{1/2} \sigma_a}} \exp{\left( -\frac{(p_a-p_{0,a})^2}{4\sigma_a^2}\right)}\\
        &\times \frac{1}{\sqrt{(2\pi)^{1/2} \sigma_{\phi}}} \exp{\left(-\frac{(p_{\phi}-p_{0,\phi})^2}{4\sigma_{\phi}^2}\right)}  \,,
        \end{split}
    \end{equation}
    where the free parameters $p_{0,a}$ and $p_{0,\phi}$ are the mean values of the Gaussian packet, and $\sigma_a$, $\sigma_{\phi}$ are their standard deviations. 
    
    We stress that, to satisfy the adiabatic condition \eqref{eq:condizderivate}, we must consider 
    \begin{equation}
        \label{eq:appliccondizpa}
        -\frac{1}{M} < p_a < \frac{1}{M} \,,
    \end{equation} 
    so we integrate the wave packet only in this interval with a proper normalization.
    
\begin{center}
    \begin{figure}[H] 
    \includegraphics[scale=0.5]{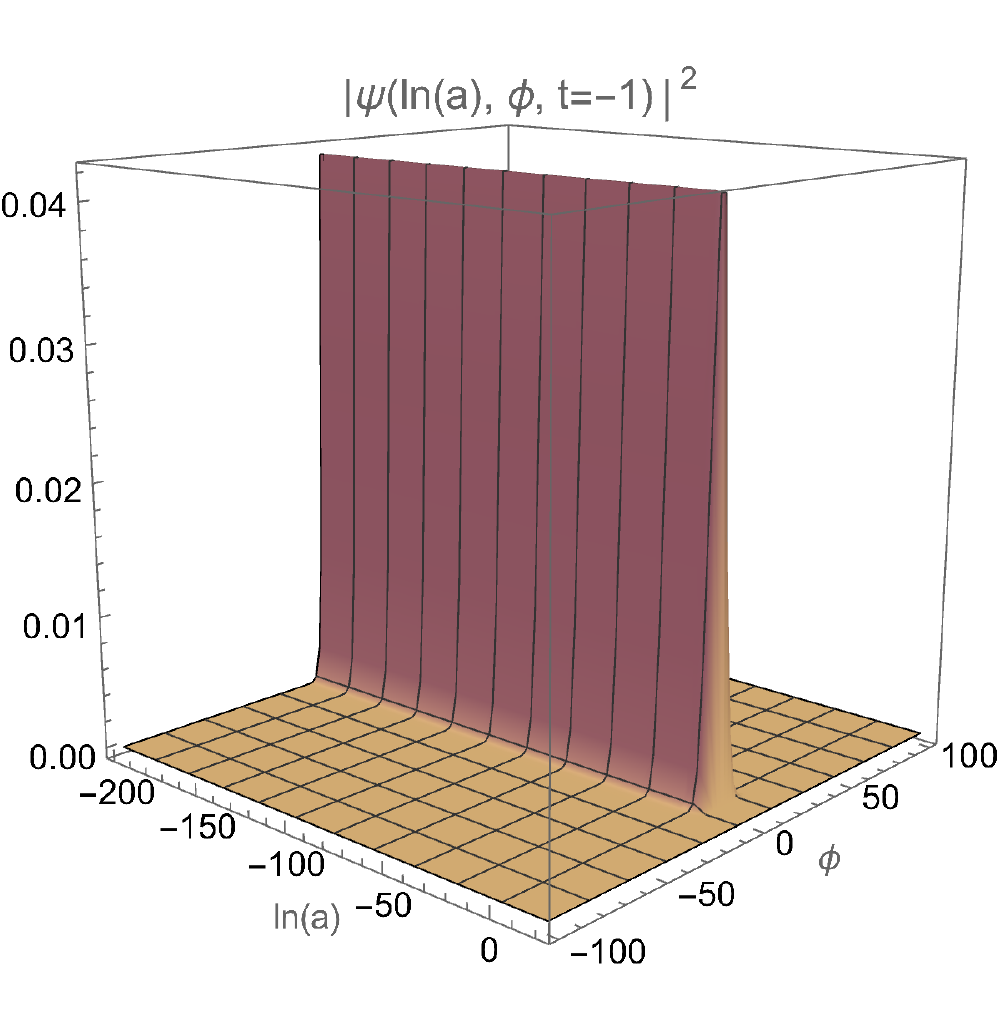}
    \includegraphics[scale=0.5]{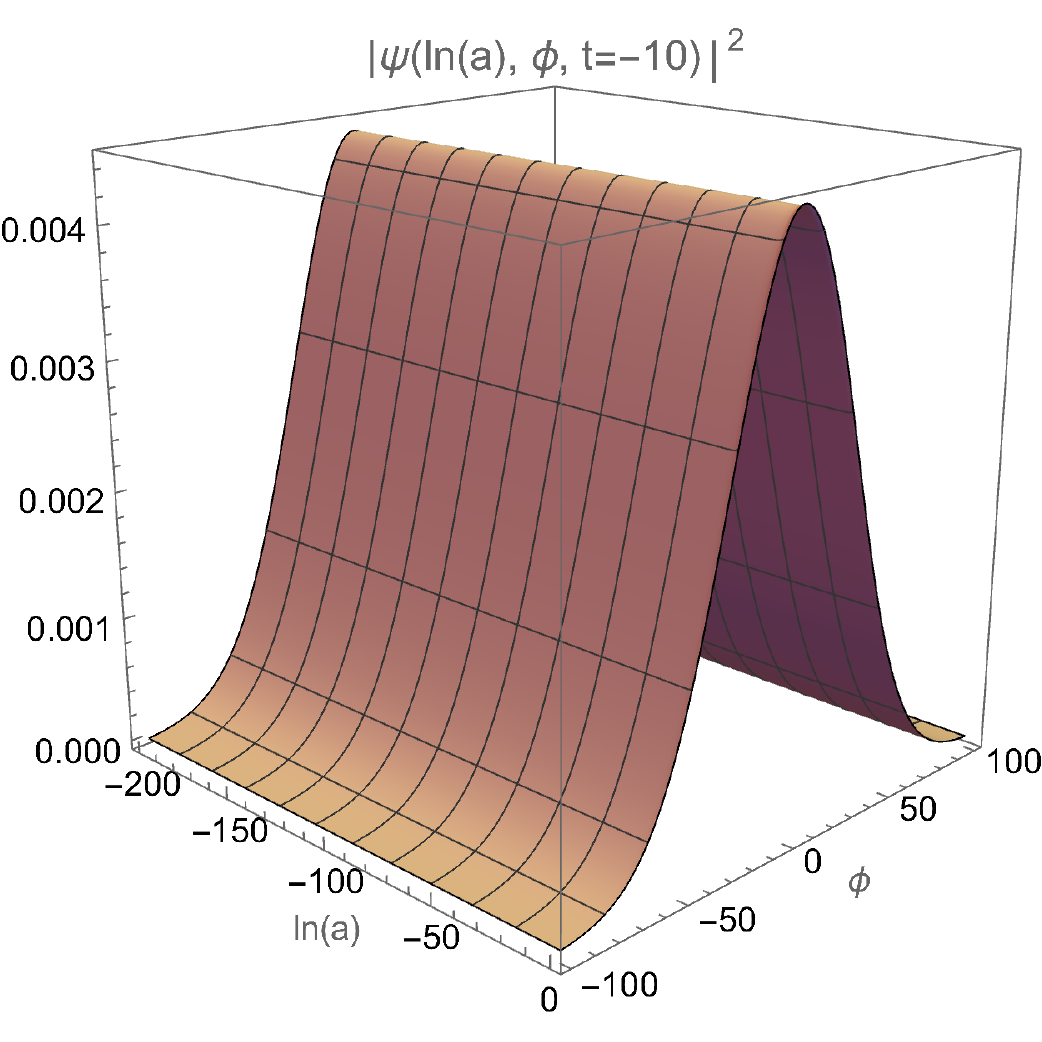}
    \includegraphics[scale=0.5]{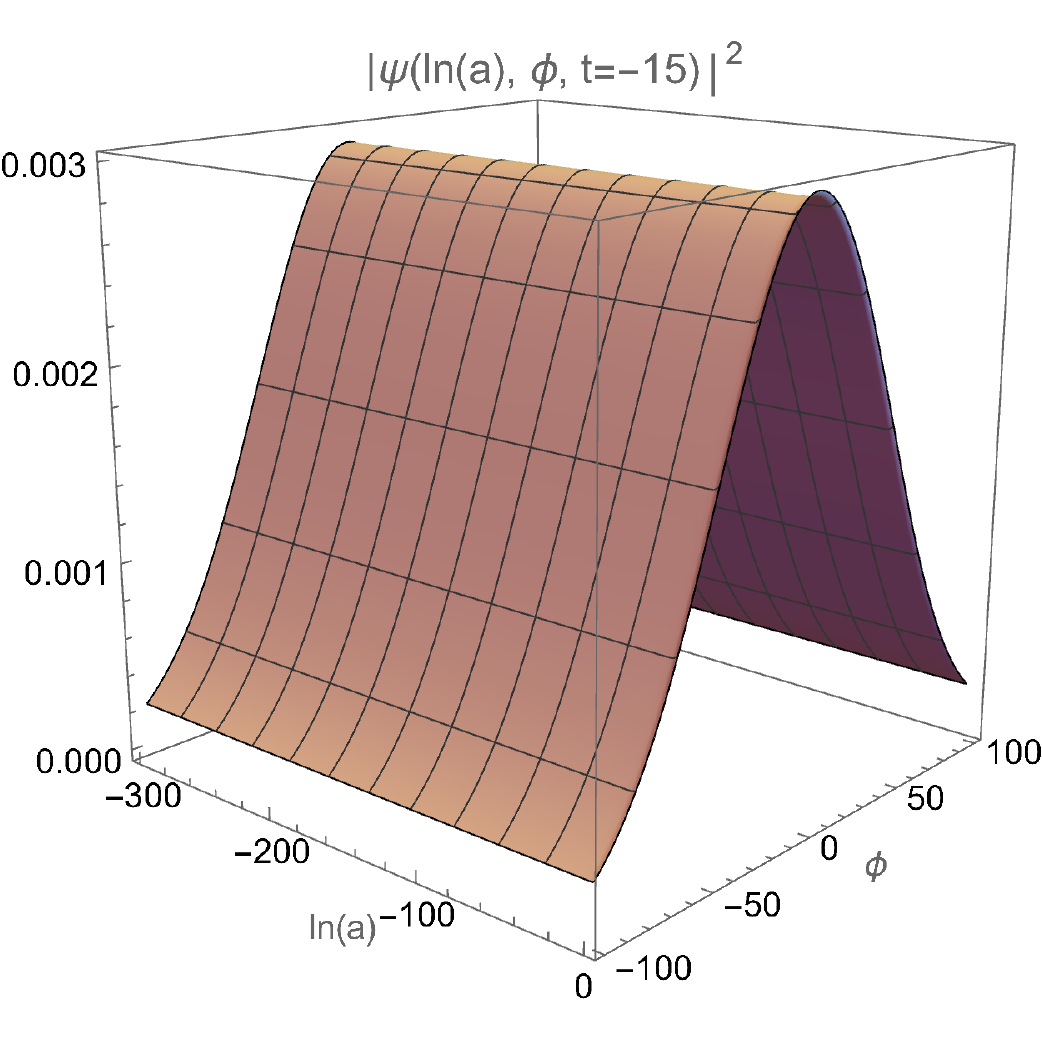}\\
    \includegraphics[scale=0.5]{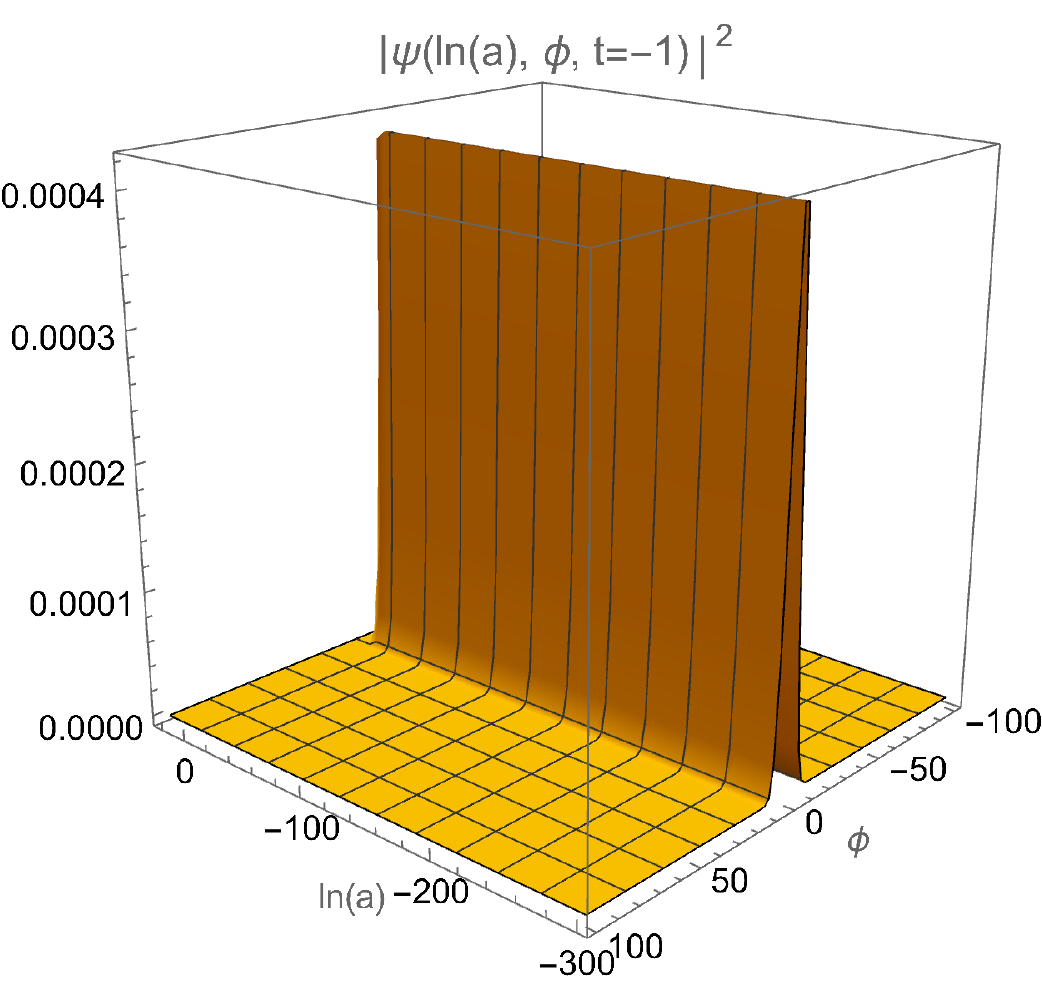}
    \includegraphics[scale=0.5]{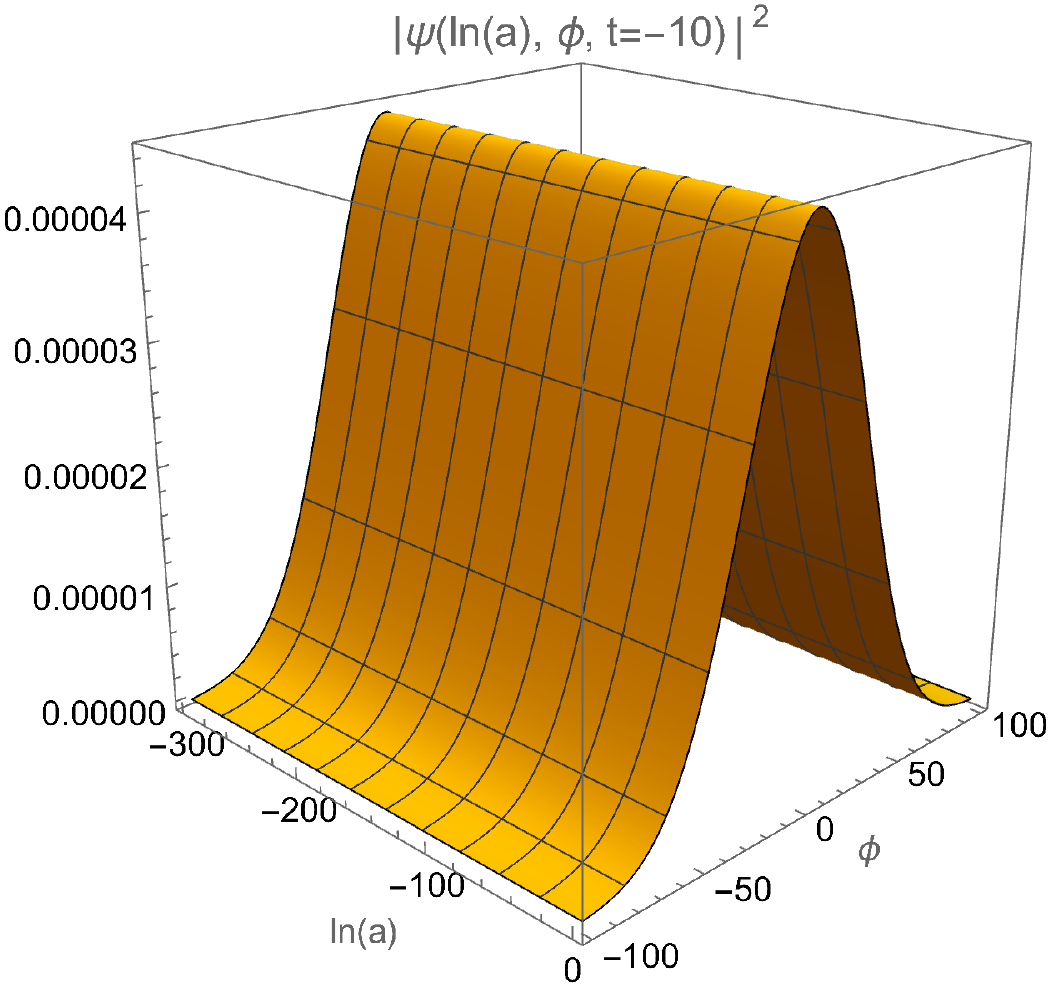}
    \includegraphics[scale=0.5]{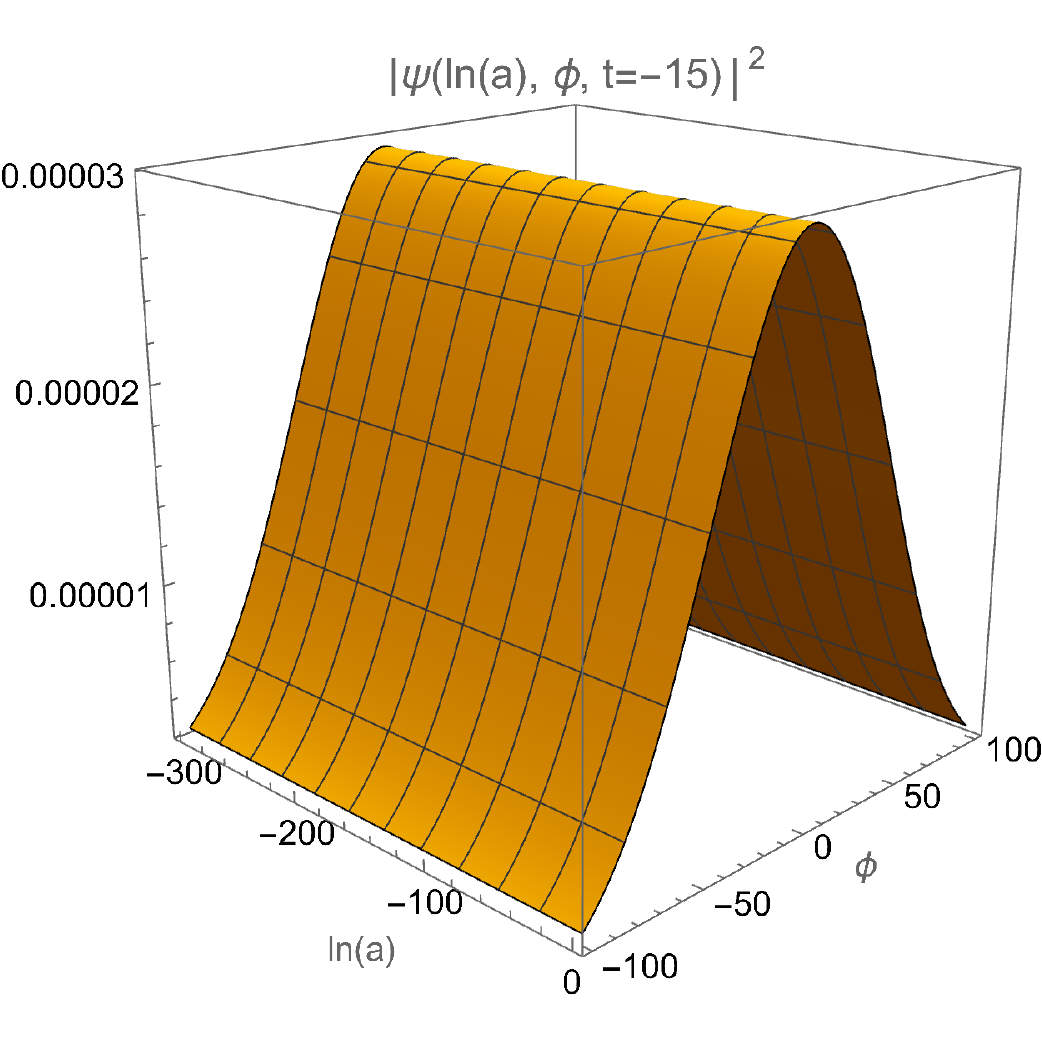}
    \caption{Here we show the probability amplitude for different values of the rescaled time $\tau$. The graphics on top correspond to the modified wave function \eqref{eq:applicsolchi1} solving the dynamics with quantum gravity corrections, while the bottom graphics show the amplitude without such quantum gravity terms for comparison. Here we have used for computation a bigger value of $M$ ($M=100$ in Planck units) with respect to the one prescribed by the theory, in order to enhance the small quantum gravity effects. }
    \label{fig:plotsModifiedChi}
    \end{figure} 
    \end{center}

    Implementing the dynamics computed above, we find that the probability amplitude for the fast quantum matter is modified as shown in Fig. \ref{fig:plotsModifiedChi}. We observe that the obtained amplitudes are almost flat (i.e. a very weak dependence) in the scale factor $a$, since the hypothesis \eqref{eq:condizderivate} requires \eqref{eq:appliccondizpa}; hence the quantum gravity effects on the system are of very small intensity as predicted by the perturbative approach. We observe that the major modification to the probability amplitude takes place as $ln(a)$ approaches zero.
    
    However, we remark that a deformation of the energy spectrum takes place, as described by the following formula:
    \begin{equation}
        \label{eq:applModifiedSpectrum}
        E = E_0 + \frac{\hbar p_a (-\tau)^{7/3}}{3 (3\Lambda)^{1/6}}\, .
    \end{equation}
    Clearly, also this corrective term lives at order $\ord{M^{-1}}$ with respect to the standard quantum field theory spectrum.
    
    \onecolumngrid
    \twocolumngrid

\section{Discussion and conclusions} \label{sec:conclusions}

    We have constructed a general paradigm to determine quantum gravity corrections to standard quantum field theory (treated in the functional representation), which adopted as a time variable the coordinates of the reference fluid emerging when the synchronous reference frame is \emph{a priori} fixed in the gravity-matter action.

    The key point of the present formulation is in the Born-Oppenheimer separation of the quantum dynamics into a ``slow component'' corresponding to the gravitational degrees of freedom only and a fast quantum subset, whose variables are identified with the matter and reference fluid degrees of freedom. Since the gravitational quantum dynamics is WKB expanded in the Planckian parameter of the theory, the request that the reference fluid belongs to the fast component implies that its presence in the Hamilton-Jacobi equation is removed; in this way, the violation of the strong energy conditions, which was originally investigated in \cite{bib:kuchar-1991} and which led us to rule out the reference fluid from the perspective of a viable time variable, no longer takes place.
    
    In our analysis, the reference fluid plays the role of a clock in the quantum dynamics of matter only, and it is, in this respect, a physical clock for the quantum subsystem.
    The main result of the proposed paradigm consisted of restoring the unitary character of the modified matter fields dynamics, differently from the analyses in \cite{bib:kiefer-1991,bib:kiefer-2018}. In this respect, however, it is worth to discuss an important phenomenological difference between our approach and previous approaches to this same problem.

    In fact, in \cite{bib:kiefer-1991,bib:vilenkin-1989,bib:bertoni-1996,bib:kiefer-2016,bib:kiefer-2018,bib:venturi-2021}, the identification of a time variable is always related to the natural label time, via the (\emph{de facto}) classical dependence of the quantum matter wave functional on the classical gravitational degrees of freedom, in turn expressed via such a label time. Thus, apart from the nontrivial question of the nonunitarity, these studies construct a quantum field theory with a label time dependence and a modified Hamiltonian operator, although dealing with quantum gravity corrections.

    Our model is instead intrinsically different: the time variable is identified among the fast coordinates, and real quantum gravity effects enter the quantum field dynamics, in the specific sense that the matter wave functional is also depending on the gravitational degrees of freedom, which are in principle quantum variables, never reduced, even in the WKB approximation, to pure classical functions of the space-time slicing coordinates.
    This conclusion is completely coherent with the idea that the gravitational background is nearly classical but never a pure classical setting and the expression ``quantum gravity corrections'' can be phenomenologically translated only by the dependence of the quantum matter evolution on an additional (weakly) quantum set of degrees of freedom.

    However, a subtle question arises here: which is the phenomenology we can infer from such a quantum gravity dependence of the quantum field theory? The question is highly nontrivial from a conceptual point of view, as it happens in almost any implementation of quantum physics to the gravitational dynamics, and especially of cosmology \cite{bib:montani-primordialcosmology,bib:cianfrani-canonicalQuantumGravity,bib:kuchar-2011}.
    However, for the cosmological model discussed here, we can make some heuristic considerations, elucidating the qualitative point of view we are proposing to interpret our results.

    Actually, we can think that, from a phenomenological point of view, we are somehow averaging on the the quantum gravity corrections to the standard quantum field theory functional. More specifically, we propose that, in the present scenario, we can reconstruct \emph{a posteriori} a modified wave functional for the matter field instead of a modified Hamiltonian. The simple way to express this phenomenological perspective corresponds to averaging the matter wave function onto the quantum gravity contributions. 
    In the present cosmological implementation, we can average the field wave functional on the semiclassical probability density for the scale factor $a$, as it emerges from the WKB approximation of the Wheeler-DeWitt. Thus, the quantum matter wave functional would read as

    \begin{equation}
    \bar{\chi}(T, \phi )\equiv \int da \,
    \lvert A\rvert ^2 (a) \, \chi (T,\phi;a) \, ,
    \label{eq:AverageChi}
    \end{equation}
    
    where by $A$ we denoted the first order quantum gravity wave function living at the next order with respect to the classical Hamilton-Jacobi function ($A= e^{S_1}$ in the previous notation).
    The dependence of the wave function $\bar{\chi}$ on the fluid variable $T$ can now be interpreted as direct dependence on the synchronous time, and it can be linked via the lapse function to a generic label time $t$.
    
    Actually, in the cosmological setting, we have provided a complete implementation of the proposed scheme, up to the first order corrections to the standard quantum field theory in the expansion with respect to the chosen Planckian parameter $M$ defined in \eqref{eq:defM}.
    We have analyzed the quantum dynamics of a homogeneous free massless scalar field living on a quasiclassical (WKB-expanded) isotropic universe. We have been able to explicitly calculate the modified wave function for the quantum scalar field dynamics. This offers a natural scheme to pursue the study of the quantum gravity correction to the spectrum of primordial density fluctuations due to the inflationary phase of the Universe. In this respect, the cosmological constant term well mimics the slow-rolling phase of the primordial Universe, and the only required ingredient to determine the spectral deformation (due to quantum gravity effects) consists of the natural intrinsic inhomogeneous dependence of the scalar field. 
    
    This generalization of the present analysis is certainly the most relevant phenomenological issue of the proposed theory, as it could also be the impact of the quantum gravity corrections on the Hawking temperature of a black hole.
    However, here we were interested in focusing attention on the morphology of the emerging quantum field theory; the cosmological analysis mainly had the role of elucidating the structure of the approximation scheme and how it must be addressed.

    We conclude by observing how the obtained modified quantum field theory coincides with the one derived in \cite{bib:maniccia-2021}, where the kinematical action has been adopted. As shown in \cite{bib:montani-2002}, also the kinematical action can be interpreted, on a classical level, as a suitable fluid, not always verifying the energy conditions.
    The equivalence of these two approaches suggests the idea that the role of the reference frame is, in quantum gravity, properly interpreted only in the spirit of a ``materialized fluid'', able to become a physical clock when it is addressed as described by fast degrees of freedom in a Born-Oppenheimer scheme.

\begin{acknowledgements}
	We acknowledge the contribution of Luca Petrolati for valuable discussion on this topic.
\end{acknowledgements}

\bibliography{ms2}

\end{document}